\documentclass[preprint2]{aastex}

\begin{document}

\title{  On the Interpretation Of the Local Dark Matter}

\author{ H. L. Helfer}
\affil{ Department of Physics and Astronomy; Laboratory for Laser 
Energetics}
\affil{University of Rochester, Rochester NY 14627}
\email{lary@astro.pas.rochester.edu}

\begin{abstract}

    The cause of the extended rotation curves of galaxies is
investigated.  It is shown  that  conventional sources and most
exotic sources  for the needed gravitational fields are implausible.
We suggest   spatial fluctuations in  a scalar field, 
 similar to the inflation field, are responsible for the gravitational fields.  These fluctuations
play the role of `dark' halos around galaxies.  They  take $\sim 10^5$ yrs to
develop and could not have been important in the early days of the
universe.  When galaxies are clustered, a $\Lambda-$term
appears naturally in this  theory.  The universe's
present energy density associated with these scalar field
variations is  $\Omega \sim 1/2-2/3$.  A possible scenario is
suggested in which  the cosmic scale factor ${\cal R}(t)$ would have
experienced a recent acceleration.   A  discussion of further observations and
theoretical work needed to resolve some ambiguities in the theory  is given.

\end{abstract}

\section { Introduction }

     This paper speculates that the local dark matter found in halos
of galaxies may actually be fluctuations in a scalar field.  The general distribution  of dark
matter has been summarized by \citet{T85,O93}.  By `local dark matter'
(LDM) we refer to the unexplained source of the extended rotation
curves, $V_{rot}(r)\ {\it vs.}\ r,$ of the Milky  Way galaxy
\citep{FT} and other galaxies.  This definition  means that the LDM
may not be the sole component of the
dark matter considered in cosmology.   But we deal
with this restricted class because the Sun is on the `flat' portion of
the Milky Way's  rotation curve and we can calculate  the LDM's 
density in our vicinity;  from this one can  derive severe constraints upon the halo's
possible composition.

       The rotation curves of  dwarf, elliptical, and spiral  galaxies have been
reviewed by \citet{CG}, \citet{ZF}, and \citet{SR}.  While a correlation  is
seen between the shapes of the rotation curves and the surface brightness and
morphology of galaxies, modeling them   requires  a mixture of baryonic matter and
LDM for which the effective density falls off as $r^{-2}$  at large $r$.

     So, rotation curves are generally observed to be
composite  with  the LDM  contribution dominating the rotation curves of normal spiral
galaxies outside their central core regions; see \citet{R80,R82,R85}.  They
may be schematized by a linear rise of the rotation velocity $V$ to a value
$V_h$ at a galactocentric distance $r=R_h$, followed by a constant rotation
velocity ($\pm \sim 10\  {\rm km\ s}^{-1}$) for $R_h <r <R_l$; here $R_l $ is set only by
observational limits of finding radiant baryonic material.  About  74\% of field
spirals show these simple  rotation curves \citep{SR}.

      We will ignore those galaxies in which large baryonic contributions to
the total mass make it difficult to extract the LDM's effect on the rotation
curve. These include: galaxies with peculiar shaped rotation curves in $r \leq
R_h$ \citet{S98}; compact bright galaxies which may have declining velocity
curves for $r > \sim 2-3R_h$ \citet{CG}; and low surface  brightness galaxies
for which $V_h $ is small $ <100\ {\rm km \ s}^{-1}$, and there is clear evidence of
large amounts of baryonic material at $r \geq 20 $ kpc, \citep{P97,QP,MRB};  we
regard these as nascent galaxies.

      Observations of the  rotation curve of the Milky Way galaxy \citep{C,M92}
show it is composite, with this typical underlying LDM distribution.
For our galaxy, adopting a solar galactocentric distance of $R_0
\simeq 7.8$ kpc \citep{FT}, one has $ R_h \simeq 5$ kpc and $V_h \simeq 200 - 210 \ {\rm
km\ s}^{-1}$. The limiting value $R_l$ for which $V=V_h$ is uncertain; it is
generally assumed to be at least 25 kpc and is more likely to be at
least 50--60 kpc and may well be $> \sim 200$ kpc \citep{FT,KL}. In some spirals, one
actually observes $R_l>60$ kpc. If the LDM is the same as the
dark halo material  used in galaxy-galaxy gravitational  lens models, then $R_l
\geq 200$ kpc is appropriate \citep{GS}.  We shall adopt $R_l>10 R_h$ in
calculations. 
     
     For other spiral galaxies, generally $R_h$ has about the same value.
However while $V_h$ often  is $\simeq 200 \ {\rm km\ s}^{-1}$,  values in the range
$ \sim 100$ to $250\ {\rm km\ s}^{-1}$ do occur. There also is a smaller class of
galaxies of interest in which the rise is more gradual, $ R_h \simeq
10$ kpc, with
the rise from $r=0$ to $r=R_h$ better represented by a convex curve, see
examples in \citet{PS}; Generally these have not been observed  much beyond
$R_h$.

    The  Milky Way schematized velocity curve, assuming spherical symmetry,
gives for $r \leq R_h, \ \langle \rho \rangle=6.4 \times 10^{-24}\ {\rm
g\ cm}^{-3}=0.086\  {\it M}_{\sun}\ {\rm pc}^{-3} = 4.0\  {\rm hydrogen \ atoms
\ cm}^{-3}.$  For $r \geq R_h$ one has $ \rho_{LDM} = {1 \over 3} \langle \rho
\rangle ( R_h/r)^2$.  At the Sun's galactocentric distance, $R_0$, one has
$\rho_0= 0.87 \times10^{-24}{\rm g\ cm}^{-3} =0.012\ {\it M}_{\sun\ }{\rm pc}^{-3}
= 0.54\ {\rm hydrogen\ atoms \ cm}^{-3}.$  Masses  of interest are: ${\it
M}(R_h)= 4.5 \times 10^{10}{\it M}_{\sun}$;  ${\it M}( r \geq R_h)=(r/R_h){\it
M}(R_h)$; and    
${\it M}(R_0)= 7.0 \times10^{10}{\it M}_{\sun}\equiv {\it
M}_G $.  For $r<R_h$ the calculated mean density may not be too useful for
representing the LDM because only a contribution for the central spheroid bulge
has been subtracted and an uncertain stellar disk contribution has been
ignored.  In this region, however, the density of the LDM must fall appreciably
below the $r^{-2}$ law ( for otherwise the horizontal portion of the velocity
curve would continue inward.) The value of $R_h$ depends somewhat
upon this uncertain baryonic disk contribution to the rotation curve for $r \le
R_0$ \citep{KG}. However, for $r>R_0$, the calculated masses and densities
are dominated by the LDM contributions ( {\it e.g.} \citet{M92}) and
using  these calculated densities and masses for representing the LDM
cannot overestimate the true values   by more than a factor $\sim 2$
if the rotation  velocity is to have  the observed value and lack of
$r-$dependence.  Similarly the value of $R_h$  may be somewhat underestimated.

   In our vicinity where $\rho \propto r^{-2}$,  the LDM column height, is $
\int_0^{\infty} n\ dz=\pi R_0 n_0 /2=2.0 \times10^{22}\ {\rm hydrogen \
atoms \ cm}^{-2}= 145{\it M}_{\sun}\ {\rm pc}^{-2}$; up to $1$ kpc the projected
surface density is $11.8\ {\it M}_{\sun}\ {\rm pc}^{-2} $.  If  the LDM were hot gas,
the emission measure would be $\int_0^{\infty} n^2 dz=\pi R_0n_0 ^2 /4= 1.8
\times 10^3 \ {\rm cm}^{-6}$ pc.

    In the next section, we show that no  population of observed astronomical
material can play the role of the LDM.  We also show that proposed
exotic sources, massive neutrinos, primordial mini-black holes, black fluids,
etc. are highly unlikely to be the LDM. 
 
     In the third  section,  a  scalar field source term is examined
which produces  gravitational accelerations similar to that observed.  The fourth
section contains  proposals for observational tests and a discussion of  some
uncertainties in the theory that still need to be resolved.

\section{ Unacceptable  LDM Sources }

\subsection{ No Interstellar Constituants}

   The observed LDM mass density for $r=R_0$ is very much higher than the observed
energy densities of magnetic field, cosmic rays, and radiant energy; these
cannot be  primary constituents of the LDM.   The LDM cannot have a
significant component of  ordinary interstellar gas, for the column densities
needed are much  higher than those observed; see pg 525 of  {\it Allen's Astrophysical
Quantities} \citep{AQ}, hereafter referred to as {\it AQ}.    A very high
temperature medium is excluded  by the low emission measures
observed in the x-ray region \citep{ MC, MS}.  One cannot make the LDM
 out of, say $\sim 5-10\  {\it M}_ G$ of He gas, without providing an explanation
as to where the missing $\sim 15-30\ {\it M}_ G$ of H, expected by observed
abundances, disappeared. Nor can the LDM be explained by large  dust particles.
By weight, dust is comprised of 'metals' with universal abundance ( by
weight) ${\it Z \le }\sim 0.02.$ For every $1\ {\it M}_G$ of LDM
attributed to dust one  must account for a missing
$50\ {\it M}_G$ of H \& He.  Consequently all particulate matter up to the size 
of Uranus and Neptune can be excluded, as well as very  small compact molecular
clouds \citep{A01}.

\subsection{  No Stellar Constituants}

     The mass distribution in the solar vicinity is given by Table 19.9
of {\it AQ}.   The only low luminosity stars that could contribute significantly to the LDM
are some low mass main sequence stars and white dwarfs/neutron stars.
Looking at edge-on galaxies, the surface brightness for any class of stars with
a constant {\it L/M} should fall off as $1/r$; this is not observed implying
only very low luminosity stars could be a main LDM constituant. Trimble ({\it
AQ,} p.530) gives as an lower limit for our galaxy, {\it M/L} $\sim 18$ to
$r=35 $ kpc; this corresponds to stars with {\it M}$ \sim 0.1{\it M}_{\sun}$ ({\it
AQ}, p.487).  But the integrated stellar  mass function up to { \it M}$ = 0.1 - 0.2\
{\it M}_{\sun}$ is only $22\% -35\%$ of the total  main sequence stellar mass
function ({\it AQ}, pg 488). Therefore, if  the LDM  of $\sim 4 \  {\it M}_G$
seen up to $r=35$ kpc is due to such low mass stars, one must  account for a
missing $12 -20 \  {\it M}_G$ of  main sequence stars associated with them
by current ideas of star formation.   Similarly, white dwarfs and neutron
stars are excluded by the same type of argument.  In their
formation at least a comparable mass of material is expelled into the
interstellar medium and it is not observed.


    Now, these arguments cannot exclude the possibility that a very old population
of stars was formed (or captured) with an initial mass function favoring objects
of  $1 -100\ {\it M}_{Jupiter}$ or that populations of  small black
holes were somehow produced  {\it ab initio}.  Normally, such proposals are
not detailed enough to discuss either the lack of expected associated matter or
the type of physics  needed to constrain the mass range.  It is also
possible   that  massive  mesons   or some other exotic particles are
present. These possibilities can be restricted by the more general
physical arguments, given in the next subsections.

\subsection{  'Dark Fluid' Hydrodynamics}

    Suppose the energy-momentum tensor of the LDM can be represented by a simple
fluid with scalar pressure, $ P =P(\rho)$.   First, assume hydrostatic
equilibrium.  Then, for the outer region $r >R_h$, using $\rho \propto
r^{-2}$,  one can  solve for an equation of
state.  One finds $P ={1 \over 2}V_h^2 \rho + P_0$ where $P_0 \propto
\rho^{V_h^2/2c^2} \cong\ $constant  , so that
for $r>R_h$ we have an isothermal sphere solution.   For finite $\rho$  at
$r=0$, the isothermal Bonnor-Ebert solutions, \citet{E55,B56}, discussed by
\citet{A01}, are appropriate .  They have too small a range in which
$ \rho \propto r^{-2}$.  Also, for our thermal velocity, $\sim V_h$, they are
 Jeans unstable for $R_L> \sim 20$ kpc and  hence not acceptable.  So,
hydrostatic equilibrium  with a simple equation of state, is
unlikely.  But, this argument would not necessarily exclude a gas in a state of partial
ionization.

      Non-hydrostatic equilibrium models can be rejected.  Postulating
centrifugally supported  ellipsoidal LDM matter runs into
another serious problem.  Suppose $V_{\phi}^2/r = (1-\xi)V_h^2/r$ for 
sufficiently large $r>R_h$; then the effective thermal velocity is reduced by a
factor $\surd \xi$. However, one would then have to explain  why the total LDM
angular momentum  increases $\propto r^2$.

    Turbulent support by means of a term ${\bf v} \cdot \nabla v_r$ is unlikely.
Mass conservation  requires $4\pi R_0^2 \rho \langle v_r \rangle  \ll  \sim 1{\it
M}_G/10^{10}$ yrs; this gives as an upper limit $ \langle v_r \rangle \sim 1\
{\rm km\  s}^{-1}.$  It would be difficult to maintain this low average radial
velocity over the entire circumference at $r=R_0$ unless $ \langle v_r^2
\rangle$ is small, say$ \sim (5\ {\rm km\ s}^{-1})^2$. Using  $ \langle v_r^2
\rangle^{1/2}$  as representative  turbulent velocity, one
would require a  characteristic turbulent length scale $< \sim 5$ pc to balance the gravitational
acceleration $V_h^2/R_0$.  Again, a mechanism for providing this small scale would be 
required.     

   Since $\rho \propto r^{-2}$ is observed, mass conservation would limit
outward (or inward) streaming of massive particles to $ \langle v_r
\rangle \sim 1\ {\rm km/ s}^{-1}.$  This is an unlikely restriction on
any hypothesized relativistic particles.

\subsection{Galaxy-Galaxy Collisions}

     Galaxy-galaxy collisions involving halos  pose difficulties for models of
the LDM involving particles not interacting with ordinary matter nor
capable of radiating.  If the particles undergo collisions among
themselves, then the halos should relax {\it adiabatically} after a
galaxy-galaxy collision, not isothermally (as is observed). If particle-particle
collisions are unimportant (as {\it e.g.} mini-black holes), then the particles'
 average velocity must be of order $V_h$ or less for them to remain bound to a galaxy.
Then if two galaxies with halos collide, one knows that the particles'
trajectories would be modified by the non-spherically symmetric gravitational
potential present during the long  galaxy collision.  After the galaxies
separate, there would be no restoring force to re-establish sphericity for the
halos, and the time evolution of the particles' distribution function would be
governed by Liouville's equation. 

      This is a good argument which is not yet conclusive  because we do
not have observational studies of halo asymmetries at large $r$.  In
principle it is possible to study the aftermaths of halo
collisions. About $5 \%$ of the galaxies are in rich clusters of galaxies
(see Bahcall {\it AQ}, p.613) and in the core regions , $\sim 1.5
h_{-1}$ Mpc radius, the average galaxy halo with   $R_l > 50$ kpc 
experiences at least one halo-halo collision in $10^{10}$ yrs with a short   collision time
$<10^9$ yrs, the collision frequency being $\propto R_l^2$.
Similarly, In  galaxy groups  and poor clusters, which contain $\sim 50 \%$
of the galaxies, the average  halo with radius $R_l >250$ kpc experiences  collisions in
$10^{10}$ yrs. In some groupings, of course such as  the Milky Way-M31
Local System   even smaller halos should show severe distortion if composed of
particles.  Field galaxies should generally keep undistorted their
primitive halo structure.

    Assuming halos show approximate spherical symmetry out to $\sim 50
$ kpc , we need conclude that  any  exotic LDM  candidate  must be able
to radiate efficiently during galaxy-galaxy collisions, either at presently
undetectable wavelengths ($\nu < 10$ Mhz) or in very low energy particle
emission ( {\it e.g.} pairs of neutrinos) to avoid adiabatic
relaxation in collisions, and that, if considered as a fluid, the equation of state
cannot be reduced  to a simple $P-\rho$ relation if  gravitational
instability is to be avoided.

\section{The Scalar Field}

    As an alternate to specifying  radiative properties and thermodynamics of
unknown  particles, we suggest  that a class of fluctuations in a scalar field
$\phi$ form gravitational potential wells into which baryonic matter may flow,
forming luminous galaxies in their center regions.  These potential
wells have gravitational mass and are the dark halos. There is no
reason to assume all halos  have luminous galaxies associated with
them, or that all  luminous galaxies  are associated with these
dark halo potential wells.

     The halo potential wells will be specified by the energy momentum
 tensor $T_{ab}$  determined  by
 $\delta (\surd (-g) {\cal  L})=-\surd (-g)(T_{ab}/2)\delta g^{ab}$  once a
Lagrangian ${\cal L}(\phi)$ is chosen.  The associated  field equation is given by
  $\delta {\cal  L}/ \delta \phi=0$. We use as a Lagrangian 
 ${\cal L}=-{1 \over 2}\alpha^2 L_a^a +\Lambda $ 
 where  $L_{ij} =\phi_{,i}\phi_{,j} +m_im_j[\phi^2(1-{1\over 2}\phi^2)-\lambda]$.
  Here  $m_i$ is an assigned timelike vector,  $m^am_a =m^2$,  and  the
coupling constant is the dimensionless $\alpha^2$, not  the usual
$\kappa =8 \pi G c^{-2}$  used for fluids. If one substitutes 
 $\Psi=\alpha \phi $, and $\Lambda =\lambda =0$, then
${\cal L}$ is the \citet{GL}  Lagrangian used in studying phase
transitions and the Higgs field and  introduced into astronomy for the
inflation field by  \citet{G81}.  The parameters $\Lambda \ \& \
\lambda $ do not affect the behavior of
$\phi$; they are introduced to allow us to shape the extent of each individual dark matter halo.
The  energy-momentum   tensor then
is  $T_{ij} =\alpha^2(L_{ij} -{1\over  2}g_{ij}L^a_a )+g_{ij}\Lambda$.  The  form of
the wave equation in flat  space is:

\begin{equation}
\label{1}
 \partial^2_{tt } \phi - \nabla^2 \phi  - m^2\phi (1-\phi^2)=0 .
\end{equation}

       Since $ T_{tt} \propto  \phi^2$, the requirement that a field theory
 has an effective mass density falling off as $r^{-2}$ at large $r $ and finite at the
 origin, basically forces the use of form of the wave equation  in
flat space to be that shown,  restricting  permissible  forms of
${\cal L}$. The wave  operator gives  the $r^{-2}$ behavior for the
effective density at large distances; the sign of the  `mass' term
$m^am_a $ is opposite  that used in the Klein-Gordon equation
 because we need solutions finite at the origin;  and a  non-linear term must be 
introduced in the potential term to limit growth.  We use a scalar
field because it is simple and it is conceptually economical  to see if a
descendent of the  inflation field can play a contemporary role.

\subsection {  Approximate Halo Solutions}

    We summarize the discussion of the flat space solutions  given
in the Appendix.  We assume that the field always has a background
of infinitesimal fluctuations, similar to those  seen in the CMB,
and  consider only solutions which could have grown from these
fluctuations $\phi(r, t=o) \sim 0$ and remain finite.  The nonlinear
cubic term has two significant functions: (1) it
limits the growth of time-dependent solutions; and (2), it may provide a strong
interaction between two sets of fluctuations. Aside from these, the term has
only a modest effect on the analytic form of the two classes of solutions found.

    The  first class of solutions are the large wavenumber modes, $k^2>m^2$.
They represent traveling waves, {\it t-waves},  for short, whose frequencies
are amplitude dependent. [ Wave packets of the t-waves are problematical; they
would act as tachyons since the group velocity $ d \omega /dk =k/ \omega$ is
greater than that of light.] These t-waves are stable, non-localizable, and
remain infinitesimal for $t >0$; for this reason they will be mostly  ignored in  this paper.

  The second class of solutions, composed of modes  with $k^2 < m^2$, do not
'travel' and can grow to finite amplitude.  With appropriate boundry
conditions, they exhibit anharmonic periodic motion, first growing
exponentially fast from a small initial value until $\phi^2 = \langle \phi^2
\rangle \le 1$ ; then de-acceleration takes place and the solutions reach their
maximum amplitudes and then decrease once more to their initial amplitudes. The rise time
from the fluctuation level seen in the cosmic microwave background, to
the maximum value is $\sim  1 -5 \times10^5$ yrs for $a^2 \sim {1 \over
3}-1$, and $m^{-1} \sim 3\  {\rm kpc}\ \sim 10^4$ yrs  (see below).They could  not have
 arisen early in the history of the universe.  They are standing waves
(or {\it s-waves}.)

      We suggest  these represent galactic halos. Because
their time derivatives make relatively small contributions to the
energy-momentum tensor, (see section A.3) time variations for the
s-waves  can usually be neglected and the steady state
solution, $k^2\approx m^2$  is
our model for a typical galactic halo.

     The  approximate steady state solution \footnote{ We use
$m_r=m_{\theta}=m_{\phi} =0$  and $t$ for $ct$ in formulae. We regard $a^2$ as small, and 
ignore the small $(\partial_t\phi)^2$ terms and terms of order $\sim
V_h^4/c^4$.   The time dependent s-wave halo solutions  are of the  approximate form:
     $ \phi=[1+h^2 \sin \Theta (h^2,t, r)]^{1 \over 2} \phi_s$ where $0 \leq
h^2(r)<1$ and $\Theta$ is an elliptic integral (See the Appendix).}
is, using  $\hat m \equiv m \surd(1-a^2)$ and equation {A6}:
\begin{equation}
\label{3a}
 \phi_s \simeq a {{\sin \hat m r}\over {\hat m r}} \approx a \  {\rm for}\
 \hat m r<\pi/2;
 \end{equation}
\begin{equation}
\label{3b}
\phi_s \simeq a {{\sin m(r-a^2\pi/4)}\over {m r}} \approx {{\sin
(mr)}\over {mr}}\ \ {\rm for}\ \hat m r > \pi.
\end{equation}
Here, $a$ is an arbitrary amplitude $a^2 \le 1$.

     One may use the steady state solution as representative  for  evaluating the
energy momentum tensor for s-waves,  using the approximations
 of equations (2) \&  (3).  For the  radially symmetric interior  Schwarzschild metric,
 $d\tau^2 = B(r) dt^2 - A(r) dr^2 -r^2 d\Omega^2$, standard relations
( {\it see, e. g.} {Weinberg (1972)}) give:
\begin{eqnarray}
\label{4}
      r A^{-1} & = & r-\int [ \kappa \rho_0+\Lambda ]   r^2\ dr 
    \nonumber \\
       \ln AB &= &   \int \ [2 \kappa \rho_0]  r\ dr \ \ +\ {\rm constant};   \nonumber \\
      f \cong  -\Gamma^r_{tt}  &=&  -(2A)^{-1}\partial_r B .
\end{eqnarray}
where
\begin{equation}
 \kappa \rho_0 ={1 \over 2} \alpha^2[(m^2\phi^2 + (\partial_r
\phi)^2)-\lambda  m^2],
\end{equation}
  and $c^2f$ is the radial  gravitation acceleration a non-relativistic
particle experiences because of this  s-wave  solution. 
  
   For the case $\Lambda=\lambda=0$, 
\begin{equation}
\label{5}
    r(1-A^{-1})  = {1 \over 2}a^2\alpha^2r[1-\sin^2mr/(mr)^2]  
\end{equation}
\begin{equation}
      f  =  -a^2\alpha^2 (2r)^{-1}[1-\sin 2mr/(2mr)] \\
 \equiv -(G/c^2) {\it M}_{halo}/r^2. \\
 \end{equation}
     For spherically symetric metrics, there are two different
definitions of the mass if $AB \ne {\rm constant}$; the first is the
volume integral of the energy density ( found in the solution for $A$)
and the second is the effective mass determining the differential acceleration of a
test body, $f$ (found in the solution for $\partial_r B$).  We give
mass formulae according to the second definition.  Outside a cluster,
where $AB=1$, the mass defintions are equivalent.

     The parameters of the theory are easily determined. The rotation
velocity,  $-(V(r)/c)^2/r$ is determined by $f$.   Since $R_h \simeq
5$ kpcs for the Milky Way, one determines $m^{-1} =3.2 $ pcs by setting $2mR_h =\pi$ .
Similarly, one finds $a_{mw}^2\alpha^2=2 (V_h/c)^2=8.8 \times 10^{-7}$; from the
discussion in  Section 3.2, we suggest  $a_{mw}^2 \approx {1 \over 3} \sim {1 \over 4} $.
This choice is also suggested by the fact that the  maximum  observed rotation
velocities of galaxies,  corresponding to $ a \simeq 1$  is less than
twice that of the Milky Way.   If  the stellar and LDM contributions to a
galaxy's rotation curve can be separated in the region  $r<R_h$,  one could
separately determine  $a^2\ \&\  \alpha^2$ from $d\ V_{rot}/ dr$ and
$V_h$   in cases  where  $a^2 \sim 1,$ since the acceleration  depends
upon $\hat m$  rather than on $m$ there.  

\subsubsection{ Isolated Physical Halos}

    The infinite  halo solution shown above  is for one halo in an otherwise
empty universe.  It  needs modification  because exceedingly small values of
 $\phi$ should not significantly contribute to the energy momentum
tensor.  There is no obvious physical interpretation for such contributions
when they produce densities  much less than  mean baryonic densities in the
neighboring intergalactic medium, or much less than the critical
cosmological density $\rho_c$.  In studying halo
stuctures, we may supress  these low background density contributions by introducing a
reference  energy level $\lambda=\lambda_0 > 0 $,
or a related limiting value $r=R_l$, by imposing other physics. One 
introduces a `cut-off` for the halo $T_{ab}$ by chosing $\lambda$ to require
that  at $r =R_l \equiv ( a/m) \lambda_0 ^{- 1/2} $ one has  $ \kappa
\rho_0 =0$ and  $ d  \ln (AB)/ dr \vert_{r=R_l} =0.$  Then, we  may
join an exterior Schwarzschild solution for $r>R_l$ to the interior
halo solution, in effect  regarding $R_l$ as the end of
the physical halo. Suppose $mr \gg 1$. The halo mass interior to $R_l$ is
\begin{equation}   
  ( G/c^2) {\it M}(R_l) =(1/6)a^2 \alpha^2  R_l, 
\end{equation} 
 The gravitational acceleration
for the interior region $r \le R_l $ is then given by 
\begin{equation}
\label{f6}
f=  -a^2\alpha^2 (2r)^{-1}[1-\sin 2mr/(2mr)] 
+r \lambda_0 \alpha^2 m^2/3. \\
\end{equation}
So that $f\to - (G/c^2) {\it M}(R_l)/R_l^2$ ,as $r \to R_l$; for  $r \ge
R_l $, one has
 $f=-\alpha^2 a^2 R_l/6r^2$. 
This represents an isolated physical halo appropriate for field galaxies. 
 Since the quantity $f$ is an observable, one should see  that the rotational
velocity does decrease in the outer parts of halos,
 $V(r) =V_h [1-{2 \over 3}(r/R_l)^2]^{1/ 2}$ for $r \leq R_l$;  again  observations could
determine $R_l$.  [Note that  we are explicitly ignoring $1/r^2$ contributions to
$f$ from any  more distant `point' sources.]

     For this case, when a halo is limited by the density of surrounding baryonic
matter,  a very crude estimate for  $R_l$ can be made. Suppose we
consider a halo galaxy  surrounded by baryonic dwarf
galaxies, each of mass $10^6- 10^7\ {\it M}_{\sun}$ with a mean spacing
$s=120-250$ kpc.  [ We used $\Omega_{dwarf}=\Omega_{baryon}/3 \approx
0.01$ \citet{BP}, a  cosmological  critical density  $\rho_c = 2.0h^2 \times \
10^{-29 }$, and  $h^2=0.5$ in making these estimates.]   For the
Milky Way-like galaxies the halo density becomes comparable to the
 dwarf galaxies' mean baryonic density when $R_l \sim s$.   Since dwarf galaxies tend
 to cluster around large galaxies, this seems to be a reasonable
order-of-magnitude guess for $R_l$ when intergalactic baryonic matter
is present.  This gives ${\it M}(s)=5\ -\ 10\ {\it M}_G$, values which
are $1/3$ of that given by equation (7) (when $\lambda =0$).  
Otherwise, an  upper limit $R_l \sim 1.6\ h^{-1}$  Mpc for Milky Way-like
field galaxies  results when the halo density becomes comparable to
$\rho_c$.   The uncertainty in specifying  $\lambda$ (or $R_l$) does not affect the
value of $f$ in the inner parts of halos.  

\subsubsection{ Clustered  Physical Halos}

          These formulae hold only when other halos are not close-by,
{\it i.e.} $R_l \ll \hat R_0$ where $\hat R_0$ is  half the mean
spacing between  halos (galaxies); we refer to this situation as case I
solutions.  For field galaxies   $\hat R_0 \sim 2.2$ Mps. In  crowded
regions such as clusters, one has another situation (case II solutions) where 
 $R_l \ge \hat R_0$;  the halo boundry must be re-examined because
the $\phi-$field may make significant contributions to
the energy density  there.  We find it necessary to introduce a
$\Lambda \neq 0$ term in order to define  reference
 background levels appreciably greater than zero.

    In the next section we note that a solution for $\phi$ can be broken up into
many effectively independent parts if their centers are far from one
another. Assuming this, one can consider that $\phi$ can be represented by
cluster of similar  dark halos  with  mean spacing  between halo
centers $r=2 \hat  R_0 \gg m^{-1}. $     Represent   each halo solution  by  a cell with a central
`bump' on top of a ` plateau', the plateau of one cell joining
smoothly onto the plateaux of the adjacent cells.  In the weak gravitational
approximation we can treat each cell separately.  Consider  two adjacent halos with central
amplitudes $a_1,a_2$.  Because we are limited by
our spherical symmetry assumption, look only at the forces and
densities  along the lines  connecting their centers. Then, at  the
join point between cells, $r=r_1$,  the individual halo densities
$\kappa \rho_{halo} \equiv (\alpha^2/2)[a_1^2r_1^{-2} -\lambda m^2] +\Lambda$
 match, providing $a_1^2r_1^{-2}=a_2^2r_2^{-2}$,  where $r_1 + r_2
\equiv 2\hat R_0$ ( and $mr_1,\ mr_2 \gg1$). We can then  regard this
location  as defining the edges of the halo bump's  interior
Schwarzschild solution,  by forcing $d \ln AB/dr=0 $ there
with the choice $\lambda m^2=a_1^2r_1^{-2}$; an exterior solution
holds for $r>r_1$.   The underlying background density (our
`plateau')  will be represented by the $\Lambda$ term; because of the local spherical symmetry
assumption $\Lambda$ is constant in a cell.  If now we choose as the
uniform background density $\Lambda  =\alpha^2 \lambda m^2/2$ we can
restore a halo's interior density $\rho_{halo}$ to be that of the
$\phi-$field, its physical value.   One has:
\begin{equation}
   f=-\alpha^2[a_1^2/2r -\lambda m^2 r/3] +\Lambda r /3 \to 0
\end{equation}
as $r \to r_1.$   For $r_1=r_2= \hat R_0$ one finds each halo bump has a mass  
 $ ( G/c^2) {\it M}(\hat R_0) =(1/6)\alpha^2a^2 \hat R_0,$
superimposed on the   mean background density level $\rho_\Lambda  = {1 \over 2} \langle
\rho_{halo}\rangle \equiv (3/8\pi) {\it M}(\hat R_0)/\hat R_0^3. $

  In each cell $\Lambda r /3$ acts like a  differential tidal
force; the  $\Lambda-$term causes the entire cluster to experience  an
expansion force. Our choice  for  $\Lambda$    requires that
the total acceleration on a test mass vanish at  cell interfaces. 

    For example, suppose we have  125 galaxies like the Milky Way, with
$\langle a^2 \rangle \alpha^2 =8.8 \times 10^{-7},$ in a cluster with radius
$R_{cl}=1.5\ h^{-1}$ Mpc. [ We ignore the possibility of dwarf baryonic
galaxies in the cluster  limiting the value of $R_l$.] Then $\hat
R_0=300 h^{-1}$ kpc.  The typical galaxy halo mass is $\sim 13h^{-1}\  M_G$
and the  total mass of the cluster dark matter is $\sim 1.7 \times
10^{14} h^{-1}\ M_{\sun}$, of which one-third is due to the background
$\Lambda-$term.  This is in the range
of rich cluster mass estimates given by N. Balcall ( in {\it AQ}).  The
agreement is surprisingly good since we have ignored halo-halo collisions,
resulting from the  galaxies' appreciable  velocities, and ignored 
density structure within the cluster.  For $N$ galaxies in a cluster,
the total cluster mass is $\propto N^{2/3}$ so that this estimate adequately
represents poor clusters as well if we use {\it e.g.} $N=8$.

 Finally, a similar argument must limit the range of applicability of the exterior
Schwarzschild solutions in case I models; the total force on a test
mass  must vanish  when $r=\hat R_0$, the mean spacing between
galaxies.   We must  introduce a $\Lambda$ term. 
Then, for $\hat R_0\ge r \ge R_l$, the force is $ f=  -a^2\alpha^2 R_l\ (6r^2)^{-1} -\Lambda r/3$
where $\Lambda={1\over2}\alpha^2a^2R_l {\hat R}_0^{-3}$.  In effect the
distinction between case I or Case II is whether or not the halo is
limited by  background  baryon density or $\phi-$field density (or by  force-balance).
   
    In  actual  cases, such as treating the Local Group of galaxies,
the convenience of assuming spherical symmetry should be replaced by the
weak field approximations for the metric tensor, requiring continuity
of the metric and its first derivatives across interfaces.  In this case, the
equations of the  interfaces between halos will be more complicated
 and $\Lambda$ and $\lambda$ will be  functions of position. A
particular halo, such as that of the Milky Way  may combine elements
of both cases I \& II.  The  rotation curve for the Milky Way should
be different in the directions toward and away from Andromeda at large
$r$ because the rotation curve  (determined by $f$) is $\Lambda -$dependent.

        The requirement $\Lambda >0$ is not dictated by the physics
 of the $\phi-$Lagrangian but is required to meet boundry conditions we imposed
 by first  subtracting off surrounding material.    As used, it is a
representation  of Mach's principle since it  explicitly appears
as a `background'  mean density induced by the proximity of other halos.

\subsection{ Close  Interactions}

    Because the theory is nonlinear, wave interactions between close galactic halos are
much more  complicated than in simple linear potential theory.  But, some simple
features are easy to see. Suppose one considers a case in which one
wants to represent $\phi$ as the sum of two components, $ \phi =\eta(r,t)+\xi (r,t)$.
 For example, one may represent the established LDM field of one galaxy and the other  the
influence of that of another passing galaxy The time of interaction between
colliding galaxies  is $\sim 10^9$ yr, much longer than the characteristic scalar
field vibration time $\sim 10^4 \ - \ 10^5$ yrs.  The coupling between the fields
induced by the cubic terms may be studied by writing equation (1) in
two parts:


\begin{equation}
 [{\partial^2}_{tt} - \nabla^2 ] \eta  =   m^2 \eta (1-3 \xi^2) 
-m^2 \eta^3 ;    \nonumber
\end{equation}
\begin{equation}
 [{\partial^2}_{tt} - \nabla^2 ] \xi  =  m^2 \xi (1- 3 \eta^2 )  -m^2 \xi^3 .
 \end{equation}
  
If $3 \xi^2,  3 \eta^2 \ll 1$, the two components act as
independent s-wave solutions, or independent halos, not affecting each
other except through gravitational interaction (see the Appendix).  If
the ineqalities do not hold, then the mixed cubic terms cause $\xi $
and $ \eta $ to act like Mathieu functions, permitting resonant
interactions to develop during the long collision time.   We shall use
a crude model.  Suppose we can time average some terms,
$\xi^2 \eta  \to \langle \xi^2\rangle \eta, \  \eta^2\xi \to \langle  \eta^2
\rangle \xi$, regarding these averages to be very slowly varying functions of
position.  Then, {\it e.g.} a quasi-static solution for $\xi$ is given by
using $m^2_{eff} =m^2(1-3\langle \eta^2\rangle)$. Consequently, if
$\langle \eta^2 \rangle >1/3$, a steady state
solution for $\xi$ is not allowed; the original dark halo of this galaxy will be
 modified and may not recover after the collision.  One needs a close collision,
centers passing within $\sim 10$ kpc, for this to occur.  In the past this may
have been more frequent; if so, one expects that an equilibrium population of
dark halos will have $\langle \xi ^2 \rangle <1/3$.  Since $m_{eff}$ may
be considerably smaller than $m$ when $\langle \eta^2 \rangle \to 1/3$ it
is possible that many of second class of galaxies, those with very long  (convex)
rise distances, $ R_h $, to be the result of  recent halo
interactions.  [Another way of getting long rise distances is to
require  $a \to 1$, since $m$ should be replaced by $\hat m$ for small $r$; see equation (2).]

   In such close halo-halo collisions, the structure of the halos are also changed 
 by the  gravitational interactions between the halos themselves and with any
 ordinary matter trapped in the halos' gravitational  wells. Section A.5 
 gives estimates of the gravitational forces involved.  It is possible that
 some dark halos have been stripped of much of their entrapped luminous material. 

      We have avoided discussing  t-waves, regarding them as infinitesimal.
But they may not be, in regions where halos collide or in earlier epochs of the
universe.  Suppose   $\xi$, represent a collection of 'strong'  t-wave fluctuations; then, in
their presence, the stability of s-wave halos will be affected. 

\section{Discussion}

   In Section 2 we eliminated all  more conventional potential sources for the
LDM.  This paper explores the idea that  dark halo galaxies are fluctuations 
in a scalar field. The s-wave  halo solutions studied are sucessful in
being necessarily spherical in their inner regions  and in producing the  form of the observed
gravitational acceleration. In its simplest form, the only parameter that can
vary from galaxy to galaxy is the square of an amplitude, $a^2<1$,  which
controls the depth of the gravitational well of the LDM.  Also, there
are parameters $m^2\  \&\  \alpha^2$ which have been determined from the
Milky Way's rotation curve;  we expect them to have these values for
all galaxy halos at the present epoch $z \approx 0$  if our Lagrangian
is complete.

Finally, there is a parameter $R_l$ which influences the gravitational
attraction in the  very outer regions of a halo. Its value may depend
upon the density of neighboring galaxies. We have guessed  that  $ R_l
>\sim  300\ $kpc for an average halo,  but   a much smaller value
might be needed  in a few galaxies \citep{CG} . 

\subsubsection{Tests and Utility Of the Theory}

           The most  important observational tests revolve around the question: What
 physics determines the values of $\alpha^2$ and $m^2$ (and of $R_l$)?  If  this
is the inflation field  then  these parameters must have varied over
large time scales and be functions of $z$.  

      If $m^2$ is constant at the present cosmic epoch, then angular
measurements of $R_h$ in spiral galaxies would provide a new cosmic distance scale.
Attempts to establish such a distance scale would provide a test
for the assumption $m^2$ is constant. One might be able to place limits on the
variation of $\alpha^2$ with time. If , {\it e.g.} $\alpha^2 \propto (1+ z)^{3/2}$
then also $V_h^2  \propto (1+ z)^{3/2} $ and the rotation curves of galaxies  at  $z
\approx 2$ would show larger amplitudes than those nearby.

  Because we may have  omitted possibly important coupling terms to matter and
 radiation, it is possible that both $\alpha^2$ and $m^2$ may have much 
 different values in the immediate neighborhood of stars, compact  clusters, or galactic centers.
 In principle, this possibility could be restricted by setting constraints upon the
rotation of the lines of apsides in double star systems and in the Sag A* -S2 system
\citep{Sch}.
      Analyzing the velocity fields of  colliding galaxies and compact
galaxy clusters provides both a test and a utility of the simple theory. The
non-linear terms should cause halo interaction forces stronger than the ones
predicted by Newtonian theory and halos should feel a stronger than expected gravitational 
attraction to ordinary matter. 

     Since baryonic matter can flow into the gravitational wells of  LDM
fluctuations  and   contribute  additional  source terms to the gravitational
fields, the rotation curves of galaxies will be composite. Use of  this scalar
theory to subtract out the LDM contribution  should allow determinations of {\it
L/M} for stars in the inner regions of galaxies.  One test is therefore whether
these determinations make sense.  Suppose there are equal opportunities  for baryonic 
matter to have settled in most of the large  LDM wells.  Chose galaxies with similar colors
for their central regions, so that
$\langle {\it L/M}\rangle $  is the same for these galaxies' central region stars.
 Then the central luminosity within $R_h \cong5$ kpc in field galaxies
would be a  function of $a^2$, measured by $V_h^2.$  This would
produce a Tully-Fisher relation but restricted to the very  inner parts
of galaxies, $ r \le R_h$.  For low luminosity galaxies, the halo gravitational
potential may actually be determinable  in the center regions. Then,  by Liouville's theorem,
one would expect to observe   star densities $\propto \exp \ [\beta \int f \ dr].$


\subsubsection{ The Equivalent Cosmological Fluid}

   In order to get a useful interpretation of  the energy momentum
tensor for  $\phi-$field fluctuations regarded as individual dark matter
halos, it was necessary to introduce reference energy and momentum
levels by subtracting from the basic $\phi$ field $T_{ab}$ a {\it  fluid}
energy-momentum  tensor, $\hat T(\lambda,\Lambda)$, where
 $\kappa \rho = {1 \over 2}\alpha^2m^2 \lambda -\Lambda, $  
$\kappa p= {1 \over 2}\alpha^2m^2 \lambda +\Lambda $, 
$U^\sigma=m^\sigma /m $, and $ \hat T(\lambda,\Lambda)^{ab}_{;b}=0.$
   In crowded fields of galaxies, $\Lambda$ represents an explicit
underlying  background field density
term used in determining the metric coefficient $A$.   It seems this
feature is unavoidable in any theory in which the source terms  are 
derived from a Lagrangian formulation.  Fluctuations in a field {\it must}
be referred to a reference level. Because the reference energy density
level in a Lagrangian normally is not included but must be specified
in  Einstein's  equation, one has a choice of either initially augmenting the Lagrangian
or of introducing $\hat T(\lambda,\Lambda)$ as an additional required
source term.  In performing the large-scale averages of source terms
required in cosmology, the intimate connection between the  $\phi$ field
$T_{ab}$ and $\hat T(\lambda,\Lambda)$  can be lost and two separate
`independent' averages may appear. One wonders if in the remote past
when  fluctuations in the Lagrangians for the weak and  strong forces were more important
contributors to the total energy density whether such   $\lambda,\ \Lambda$ terms
contributed significantly to the early cosmological $\Lambda$ term. 

      A {\it fluid} representation of the energy momentum tensor for a particular spherical s-wave
 solution  when $\lambda=\Lambda=0$  is not rewarding because $T_{xx},
T_{yy},  T_{zz}$  are not equal and vary from place to place; it is
only by averaging over the surface of an entire sphere $ r=constant$
that a pressure can be defined: $3\bar p \equiv \langle T_{xx}\rangle
+\langle T_{yy}\rangle +\langle T_{zz}\rangle =  \langle T_{tt}\rangle \equiv \bar \rho.$ 
The same relation holds for averaged t-waves. However, for a cosmological
fluid, one consisting of a great many individual s-wave regions  in a
unit volume,  one may not adopt the  the same equation of state
used for representing light, $\langle p\rangle=\langle \rho \rangle/3$,
because in this case, in general  $\lambda,\Lambda \neq 0.$  I am skeptical that
spacial averaging can be dismissed as a trivial problem, for the link
between a source field and (some of) the cosmological constant
$\Lambda-$term can easily be lost.
  The conservation laws $(T^{ab}-{\hat T}^{ab})_{;b} =0$ necessarily involve   $\lambda\ 
\&\ \Lambda $, modifying the definition of the effective density
and pressure.  Therefore, the cosmic time evolution of the s-wave field
density fluctuations need not be $\propto {\cal R}(t)^{-4}$,where ${\cal  R}$ is the cosmic scale
factor, because the  distribution of the halos themselves must be considered.

   Suppose we assume we can replace  galaxies by  halos, each of the same mass,
 ${\it M} =(c^2/6G) \langle a^2 \rangle \alpha^2R$, (see
equation (9),where $R$ is the effective maximum size of the representative s-wave
 halo. Suppose they were equidistant from one another with a number
density $\psi$.  Assume that in the recent past, $\alpha$ was constant and
halos were neither destroyed or created.  If $R$ were an assigned
multiple of $m^{-1}$, the mass of each halo decouples from the general
cosmic expansion and the LDM contribution to the cosmic mass density
would scale as  ${\it M}\psi \propto {\cal R}(t)^{-3}$ .  For case II
solutions,when the  halo mass is $\propto R_0=\psi^{-1/3}/2$,  the LDM halo
contribution and the $\Lambda$ term would scale as  ${\cal  R}(t)^{-2}$.

     The present minimum value for $\psi$ is  probably a mass (or luminosity) weighted
Schechter function (see {\it AQ}, p.581), $\psi =0.014h^3$ halos ${\rm
 Mpc}^{-3} $. Then, for case II,  using $R_0=\psi^{-1/3}/2, $ the values of $a^2\alpha^2$
found for the Milky Way galaxy and, ${\it M}(R_0)\psi+\rho_\Lambda
\equiv \Omega_{LDM}\rho_c$, one finds $\Omega_{LDM} \sim 0.45,$ independent
of $h$, with $\rho_{\Lambda} \sim 0.15 \rho_c$. Since there  can be
halos not associated with luminous galaxies, it is reasonable to consider $\psi$  for
halos is $\sim 1.5$  larger than that used; then the  values of $\Omega$
would then be $4/3$ larger. These are conservative estimates, for
suppose we just guess that each galaxy has a halo mass ten times its
luminous baryon mass, then $\Omega_{LDM} \sim 10 \Omega_{baryon}
\simeq 0.3 $.  The possibility  that  $\Omega_{LDM} >\Omega_{baryon}+\Omega_{other} $
is appreciable and  $\rho_\Lambda$ from clustering LDM cannot be
ignored.  If so, the current  cosmic  expansion rate is then determined
by the LDM with  $ {\cal R}(t) \propto t$.

    This estimate permits us to point out an interesting scenario.  Suppose we
take at present $\Omega_{other} \sim 1/3\ ( \cong  (1/2)\Omega_{LDM})$, scaling in time as
ordinary matter, $\propto {\cal R}(t)^{-3}$.  Then going back in time to
when  ${\cal R} =1/3$ one sees the reverse would have been true, $\Omega_{other}
\sim \ 2\Omega_{LDM}$ and then the scale factor would have had a different time dependence,
 $ {\cal R}(t) \propto t^{2/3}$. This would mean that in the time
between $z=2$ and the present, the universe would have been observed
to experience  an accelerated growth rate. Such an acceleration may have been
observed.  [For a discussion of the observations of Perlmutter, Riess
and their very many collaborators see \citet{P03}.] The argument can
be inverted: If these observations hold up, it would be reasonable to conclude 
 $\Omega_{other} \propto {\cal R}(t)^{-3}$.

\section{  Summary of Theoretical Uncertainties}

 What we have found is that the $\phi-$field variations, as presented,  can explain features of
 observed galaxy rotation curves.    Coupling to  other matter fields
is not presently needed  since a non-linear theory  can produce its
own source term;  however, future observations could  easily require inclusion of such terms. 

       There are some computational  investigations  needed to verify
 that the  time-varying s-wave solutions
will settle down quickly  enough to be represented by the  steady
state solutions. Also,   the fluctuation spectrum in physics is
represented by the action of  a stochastic force on the
RHS of equation (1).  In order to specify the cosmology of the
$\phi-$field,  the time evolution of this force is needed.    At
the very least it has to be shown that the evolution of $\phi$ variations from
the observed  spectrum of fluctuations seen in the cosmic microwave
background   can produce  the presently  observed galaxy clustering.
In investigating this problem, one would effectively determine the
time dependence of the $\phi-$field back to the time of photon decoupling.

    Because of the uncertanties in knowing how $\alpha^2 \ \& \
m^2$ varied with time, and what role should be ascribed the t-waves in
the very distant past, we simply do not know whether the scalar field
proposed  here is  compatable with any of the many scalar theory
suggestions for `dark matter' or `dark energy'   which have already
been proposed. Conventionally one sets for the spacially averaged
scalar field for the dark matter a fluid representation,
 $p= (\partial_t \phi)^2/2-V(\phi) - \langle (\nabla \phi)^2\rangle/6$
 and $  \rho= (\partial_t \phi)^2/2 +  V(\phi) -\langle (\nabla \phi)^2\rangle/2 $,
 with the spatial derivative terms normally  ignored. [ See \citet{KT}
and comprehensive reviews of dark matter theory  by
\citet{PR03} \&  \citet{Be03}.]   We require that the averaged spatial derivatives 
be included.   For us, the  cosmological average for $p$ is the same,
 (substituting $( \partial_r \phi)^2$ for
$(\nabla \phi)^2 $ and using $V(\phi) =- {1\over2}m^2\phi^2(1-\phi^2/2)$), but
 $\rho$ is different $  \rho= (\partial_t \phi)^2/2 -  V(\phi) +
\langle( \partial_r \phi)^2\rangle/2 $, even when the necessary terms
$\lambda \ \&\ \Lambda$ are ignored.  The two forms are not
compatable, even for t-waves. [ The differences arise from the fact we
used in the Lagrangian a factor $m^am_a$ rather than $m^2$; one gets a
different energy-momentum tensor from the conventional one when the
variation $\delta {\cal L}/ \delta g_{ab}$ is performed.  We prefer
our form of the Lagrangian because we do not get basic
changes in the algebraic form of the Lagrangian if we consider simple
transformations  in $g_{ab}$ like $g_{ab} \to -g_{ab}$.]


\appendix

\section{APPENDIX}

\subsection{Representations of the s-Wave Solutions}

    We consider the flat space field equation (1) with $A=B=1$, requiring
solutions be initially infinitesimal.  There are two classes of solutions.
Consider  initially  {\it local} mode excitations, $\phi  \propto \exp i (\omega t
+k\cdot r)$; then the dispersion relation for the modes become:
\begin{equation}
\label{2}
\omega^2 = k^2 -m^2(1-\langle \phi ^{\ast} \phi \rangle).
\end{equation}
For $\omega^2 >0$ we have t-wave solutions which remain infinitesimal.  The
second class, s-wave solutions, correspond to $\omega^2  <0$;  they
are non-traveling waves and may grow to finite amplitude, their smallest
values being determined by boundry conditions.  The Fourier transforms 
$\tilde \phi (k,t)$ for the s-wave solutions must have the feature
that  $\tilde \phi (k,t) \sim 0$ for $k^2 > m^2$ (since
these Fourier modes always remain infinitesimal.)  Consider the rest frame 
of a point treated as the origin. This restriction implies that $\phi$ must fall
off rapidly for $r > \pi/2m$  and is effectively large  only in the small
central region.
 
      For simplicity we consider $\phi$ to be a real field.  All s-wave solutions  finite at 
the origin, can be written as $\hat a(t) + \phi(r,t)$.  Then
 \begin{equation}
 \label  {A1a}
      \nabla^2 \phi = - m^2\phi (1-{\hat f}^2),
 \end{equation}
 where $\hat f(r,t)^2\le 1$ is a slowly varying bounded function [ This follows from
taking the Fourier transform of $\nabla^2 \phi$ and applying the mean value
theorem.]  All s-wave solutions must be of this form including the
steady state solutions, $ {\hat f}^2 = \phi^2$.  For solutions $\phi \rightarrow 0$ as
 $r \rightarrow \infty$, we require $\hat a=0$. The time dependence is then  given by:
\begin{equation}
\label {A1b}
\partial^2_{tt} \phi =m^2 \phi ({\hat f}^2 -\phi^2).
\end{equation}
    With suitable boundry conditions, some solutions are stationary
and some oscillate\footnote{The boundry condition at  $t=0$ is that
$(\partial_t \phi)^2$  is  small,
satisfying   equation (A7) with $0<a_0^2< \langle { \hat f}^2
\rangle.$ If this condition is not met, the solutions will
exponentially decay.} with ${\hat f}^2 $ representing a `mean' value of
$\phi^2$. 

   We need consider only those solutions for which, at large $t$,  the maximum values of
$\phi^2$  decrease as $r$ increases;  then ${\hat f}^2$ must also
decrease with $r$ because it is bounded by the maxima and minima of
$\phi^2$.  When the oscillations are of low amplitude, one finds
$\hat f \approx \phi_s$, one of the steady  state solutions discussed below. More
generally, we shall assume that in time both $ \langle \hat f^2\rangle$ and $\phi^2$ 
 approach one of the steady state values $\phi_s^2$. Since the
frequency $\sim m \surd {\hat f}^2$ also decreases with $r$, the
time variation of the  solutions  discussed below, is effectively confined to the
central regions  and the steady-state solution adequently represents
the outer regions. 

 \subsubsection{Radial Symmetry}
 
      Only spherically symmetric solutions can dominate in an inner
radial region, since the non-spherical modes cannot satisfy the restriction
$k^2_{transverse} =\ell (\ell +1)/r^2 <m^2$ until $r$ is large;  {\it e.g.} for
$m \sim 3$ kpc, the $\ell=2$ mode cannot  possibly be significant until $r >\sim
7$ kpc.   If  many other isolated halo regions of s-wave solutions exist they
will cause nonradial interactions in each other's outer regions.  Considering
the  case of all regions being of the same minimal size; then $\ell =6$ is
appropriate  for regions in contact and each  would need a minimum
 core  radius $>18$ kpc for such interactions to be represented by s-waves;
 if this restriction is not true t-waves will be generated by the interaction.

\subsection{ The Steady- State Halo Solution}

     Equation( A2), with ${\hat f}^2=\phi^2$,  may be replaced by an integral equation using a
Greens' function and regarding the nonlinear term as a source.  In the
general case, the  formal steady state  solution, can be represented
by a convergent series expansion in the amplitude $a <1$ at $r=0$.
One has $\phi_s = a(\phi_0 + a^2\phi_1+a^4\phi_2 +\dots)$, where
$\phi_0=\sin mr/mr$  (plus angular terms.)  for an isolated halo.
 The $\phi_n$ can be related to one another  by {\it e.g.}

\begin{equation}
\label {A2}
             \phi_1({\bf  r}) = {1 \over {4\pi}} m^2\int {{\cos m \vert
{\bf r-\acute r}\vert}\over { \vert {\bf r-\acute r}\vert}}\phi_0^3 ({\bf
\acute r})d^3{\bf \acute r}.
\end{equation}.
  
    Alternatively, a static (or time averaged)  spherically symmetric real
solution can be obtained by sucessive approximations.  Put  $ \phi_n = (f_n/r)
\sin g_n$; then
\begin{equation}
\label{A3}
    g_n =\int [m^2(1-\phi^2_{n-1}) +\partial_{rr}^2 f_{n-1}]^{1/2}\ dr,
\end{equation}
where  $f_n^2=(a/m) \partial_r g _n)$ is required as a constraint.To lowest order
we use $  \phi_0 = (a/mr)\ {\rm sin}\ mr $ and $\partial_{rr}^2 f =0$.
Succesive approximations amount  to an expansion in the amplitude $a$ at $r=0$.
Since it is probable that $a^2 < 1/ 3$, the next approximation should be
sufficient:
\begin{equation}
\label{A4}
 \phi \approx  \phi_1 =(a/mr)[(1-a^2)(1-\phi_0^2)]^{-1/ 4}\ {\rm sin} \int m
 [1-5\phi_0^2 /6]^{1/2}\ dr \simeq (a/mr\surd(1-a^2)) \sin m\int[1-\phi_0^2]^{1/2}\ dr.
\end{equation}

\subsection{ Time dependent Halo Solutions}

  There is a class of time dependent halo solutions which are related to the 
steady state solution, permitting us to estimate the time it takes the s-waves 
to develop. In equation (A2), for some time interval,  replace $\hat f^2$ by a time averaged
 value, $\langle \hat f^2 \rangle$, to simplify the time dependence; then equation (A2)
can be integrated:
\begin{equation}
\label {A5}
  (\partial_t  \phi)^2 = {1 \over 2}m^2 \langle \hat f^2 \rangle
  (\phi^2-a_0^2)(a_1^2-\phi^2),
\end{equation}
   where $a_0^2(r)$ is the value of $\phi ^2(r)$ at the lower turning point
and $a_1^2(r) \equiv 2 \langle \hat f^2 \rangle -a_0^2(r)$ is its value at the
upper turning point.  
 Using the notation $a_0^2 \equiv \langle \hat
f^2\rangle(1-h^2(r))$ for defining $h^2(r)$, the solution may be written as:
\begin{equation}
\label {A6a}
\phi^2 = \langle \hat f^2 \rangle H \  {\rm where} \ H \equiv [1 +h^2(r) \sin
\theta],
\end{equation}
 with the time dependence given by the standard elliptic integral
\begin{equation}
\label {A6b}
 [ 2m^2\langle f^2 \rangle]^{1 \over 2} (t-t_0(r))=
       \int _{-\pi/2}^\theta H^{-1/ 2} d \theta
\end{equation}
For small values of $h^2$ this is basically Kepler's equation relating the mean
and eccentric anomalies.  For   the steady state solution. $h^2=0.$

     A solution still requires a value of $\langle \hat f^2\rangle $.  One may regard equations
(2) \& (7) as defining $\langle \hat f^2\rangle$ in terms of $\phi^2$, by sucessive substitutions.
If this value is substituted into equation (A3), one can solve to get the actual
form of $\phi (r,t)$.\footnote{ The formal procedure is to use the specification of $\phi(r,t=0)$ in
equation (A2) to solve for $\hat f(r,0)$; using this and the
specification of $\partial_t \phi (r,t=0)$ in equation (A7), one calculates
$a_0^2(r)$.  The integration, equation (A9),  then  gives the
time evolution in the vicinity of $t=0$.  Because the time varying amplitude
so obtained is $r-$dependent, $\phi (r, \delta t)$ will have a different
$r-$dependence than $\phi(r,0)$; this causes $\hat f(r,\delta t)$ to change, 
requiring the process to be iterated for the next time step.  While $\hat f$ is bounded,
it too will oscillate with time. Equation (A3) forces mode-mixing
and some generation of t-waves which travel away from the
halo. This produces some damping of the time dependent s-wave
solution.   The use of $\langle \hat f^2 (r)\rangle$ permits a
reasonable estimate of the rise time and of the value of $\langle
\phi(r)^2 \rangle$, }  An adequate  approximation, for $h^2 \simeq 1$, is to
ignore the spacial variation of $H$ and replace $\phi$ by $\phi_s \surd H$, where
$\phi_s$ now is the (approximate) steady state equation given in the form of
equation (A6):
\begin{equation}
\label {A7}
         \phi \simeq \surd H \cdot (a/m) \sin \int m (1-a^2 \phi_0^2)^{1/2}dr.
\end{equation}

   For evaluating components of the energy momentum tensor we note time derivatives of 
s-wave solutions normally can be ignored since ${\dot \phi}^2
\sim m^2\langle {\hat f}^2 \rangle \phi^2 < m^2 \phi^2 \sim( \partial_r \phi)^2$.

\subsection {The Rise Time}

     We now consider the rise time of these s-waves. From the rotation curves,
we estimate  $m ^{-1}\sim    3\ {\rm kpc}$ so that the time scale  is measured in
units of $m^{-1} \sim 10^4$ yr. For the case $a_0^2 \ll \hat f^2$,
the rise  time, $t_1$, from $a_0 $ to $a_1$ is then approximately,
\begin{equation}
   m \hat f t_1 \approx \ln [2 \surd 2 \hat f/ a_0].
\end{equation}
The initial variations in $a_0^2 \propto T_{tt}$ are assumed to be $\sim 1
\times 10^{-4}, $ equal to the flux density variations seen in the cosmic
microwave background.  Our solutions are limited to $r<R_l$ where
$\phi(R_l)^2 \sim 10^{-4}$, for the solutions cannot be carried out  to values
less than the fluctuation levels from which they arose.  This determines
$ R_l\approx 100m^{-1}\surd a^2 \sim 170$ kpc for $a^2={1 \over 3}$ and rise times
$\sim 1 - 5 \times 10^5$ yr for the central region $mr < 10$. Therefore
the characteristic times of oscillation and of growth to full amplitude
are short compared to present galactic internal dynamic time scales,
but comparable to the age of the universe at the time of photon decoupling.

\subsection{ Curvature  and  Restricted Two-Body Problems}
     For a Schwarzschild metric, the wave equation is:
\begin{equation}
\label{1a}
 0=\partial^2_{tt} \phi - (B/Ar^2)\partial_r(r^2\partial_r \phi)
 -B\nabla^2_2 \phi  - m^2\phi (1-\phi^2)   \nonumber \\
    +{1 \over 2} [\partial_t \phi \  \partial_t \ln  (A/B)  - \partial_r  \phi \ \partial_r (B/A)],
 \end{equation}
where
\begin{equation}
 \nabla^2_2  \phi \equiv (r^2 \sin^2\theta)^{-1}[\sin \theta \partial_\theta ( \sin \theta
\partial_\theta \phi)+\partial^2_{\varphi \varphi}\phi].   \nonumber
\end{equation}
    For the weak gravitational fields considered in this paper ,the coefficients
     $A,B \sim 1$ plus terms of order $(V_h/c)^2$ and are
static. The term $\partial_t \phi \ \partial_t (B/A)$, for a
time-dependent halo, produces damping with a characteristic inverse
time $\sim (V_h/c)^2 m $ and can be neglected. The term $\partial_r
\phi \ \partial_ r (B/A)$ is of  order $\sim (V_h/c)^2 r^{-1}
\partial_r\phi $) and produces a very slight distortion of  s-wave solutions
centered at the origin.  So, neglecting terms of order $(V_h/c)^2$ the flat space 
wave equation is sufficient  to evaluate contributions to the energy momentum tensor. 
in  equations (4)-(6).

   However,  the gradient term, $\propto \partial_r (B/A),$ does  determine the motion
 of the centroid  of a distant small `test' halo, one  which makes a negligible
 contribution to the  energy-momentum tensor.  For suppose we change the independent
 variables in the field equation $ (t, {\bf x}) \to (t,{\bf w})$, the local coodinates 
centered on the test halo,   $ {\bf x}={\bf u (t)} +{\bf w}$, and as before, set
  $A,B \approx 1.$  In terms of the new independent variables, the wave equation
 becomes the normal one for flat space,  $\partial^2_{tt} - \nabla \phi -
m^2 \phi(1-\phi^2) +\dots =0,$ providing one sets
\begin{equation}
 \ddot {\bf u} = - {1\over 2}{{d\ (B/A)}\over {d\ r}}|_{r= u}\hat {\bf r},
\end{equation}
 Here $u =|{\bf u }|$,  two very small terms, of order $\dot u^2m^2\phi$ and 
 $\dot u m \omega \phi$ have been neglected, and we have used a Taylor's expansion
 of ${d\ (B/A)}\over {d\ r},$ assuming $u \gg w$.  This is an equation of
 motion for  the centroid of the distant test halo in the field of a
central halo or star.   [ The next order term in  the expansion  gives
the tidal force on the test halo.]

   If the central object generating the metric is an ordinary baryonic mass or black
 hole, one has $AB=1$ and $ A^{-1}= -2G{\it M}/r$; consequently the test halo
experiences {\it twice} the gravitational attraction that a test mass would feel.
If the central object is a s-wave halo, and $w \gg m$ then $A \approx$ constant, since
${\it M} \propto r$;  the distant test halo experiences the same
acceleration $f$  a text mass experiences (see equation 76)).  Since
halo-halo interactions are  wave phenomena; the halos do not really
make  separate contributions to the energy-momentum tensor.  In close
 massive halo-halo interactions, the energy-momentum tensor will show interference
 phenomena and the wave equation, because of the non-linear terms,will also contain
interference terms. Consequently interactions between close massive halos will not be as 
simple as those involving `test' halos.

\end{document}